\documentclass{article}
\usepackage{spconf,amsmath,graphicx}
\usepackage{hyperref}
\usepackage{xcolor}
\usepackage{multirow}
\usepackage{balance}

\ninept
\title{Towards Controllable Audio Texture Morphing}
\name{Chitralekha Gupta$^{*\dagger}$, Purnima Kamath$^{*\dagger}$, Yize Wei$^{\dagger}$, Zhuoyao Li$^{\dagger}$, Suranga Nanayakkara$^{\dagger}$, Lonce Wyse$^{*\ddagger}$
\address{$^{\dagger}$National University of Singapore, Singapore\\ $^{\ddagger}$Universitat Pompeu Fabra, Barcelona, Spain
\\ \small{$*$equal contribution}}
}
\begin{document}
%
\maketitle
\begin{abstract}
In this paper, we propose a data-driven approach to train a Generative Adversarial Network (GAN) conditioned on ``soft-labels'' distilled from the penultimate layer of an audio classifier trained on a target set of audio texture classes. We demonstrate that interpolation between such conditions or control vectors provide smooth morphing between the generated audio textures, and show similar or better audio texture morphing capability compared to the state-of-the-art methods. The proposed approach results in a well-organized latent space that generates novel audio outputs while remaining consistent with the semantics of the conditioning parameters. This is a step towards a general data-driven approach to designing generative audio models with customized controls capable of traversing out-of-distribution regions for novel sound synthesis.
\end{abstract}
\vspace{-0.2cm}
\section{Introduction}
\label{sec:intro}
\vspace{-0.2cm}
Sound morphing encompasses a set of models with the goal of producing gradual transformations between sounds \cite{caetano2011morphing}. Sound morphing is useful in applications of sound design including music compositions, video games, and sound synthesizers~\cite{kazazis2016sound}. Although there is a lack of consensus in the literature about the exact definition of sound morphing \cite{caetano2011morphing,kazazis2016sound}, there are certain characteristics of sound morphs that are commonly agreed upon. For example, the morphing transformation between two sounds is expected to produce perceptually intermediate results that should fuse into a single perceptual source that resembles both sounds at the same time \cite{caetano2011morphing, slaney1996automatic,kazazis2016sound}.

We focus on audio textures, a rich class of sounds in which certain parameters remain stationary over time \cite{mcdermott2011sound} despite statistical variation within the sound. For example, the sound of wind at a certain strength or the sound of tapping at a certain rate. Sounds with specifically varying spectro-temporal envelopes such as a single footstep, speech, or music do not fall under this definition of audio textures. Automatic audio texture synthesis is an active area of research \cite{mcdermott2011sound,antognini2019audio,gupta2021signal} that has applications in sound design and Foley synthesis systems \cite{choi2022proposal}.  

Many studies have explored morphing between musical instrument timbres \cite{kazazis2016sound, caetano2019morphing, caetano2011morphing} or voice timbres \cite{slaney1996automatic,ezzat2005morphing} using various signal processing techniques, however there have been limited studies on audio texture morphing. Morphing between two pitched musical instruments or two voiced phonemes is typically achieved through signal processing techniques such as interpolation between the coefficients of a source-filter model representation of the two sounds \cite{slaney1996automatic}, or interpolation between the harmonic components of a sinusoidal model representations of the two sounds \cite{kazazis2016sound}. Such methods have an underlying requirement that the two sounds are pitched, such as musical instruments or voiced utterances, therefore applicability of such techniques to non-pitched audio texture sounds is limited. Moreover, linear interpolation between parameters may not result in perceptually linear interpolation between the sounds \cite{caetano2011morphing}.

The goal of parametric audio texture synthesis is to generate novel sounds with descriptive parameters that match those of a target texture. McDermott et al.~\cite{mcdermott2011sound} developed a set of statistics based on a cochlear model to describe the perceptually relevant aspects of a given audio texture. Recent works~\cite{ulyanov2016audio, antognini2019audio, caracalla2020sound} have adapted the seminal work on image style transfer~\cite{gatys2015texture} for audio texture synthesis, where hand-crafted statistics are replaced with Gram matrix statistics computed as the correlation between feature activations to represent style. 
Though this method of audio style transfer produces interesting combinations of the sounds, there is no control of semantic style or content features other than through the data examples provided.

Recently, parametrically controllable audio synthesis has been used to help organize the latent space of the GAN and Variational Autoencoder (VAE) independent of the control parameters. Luo et al.~\cite{luo2019learning} learn latent distributions using VAEs to separately control the pitch and timbre of musical instrument sounds. Engel et al.~\cite{engel2017neural} conditioned an autoregressive model to interpolate between musical instruments to generate new sounds. The GANSynth architecture \cite{engel2019gansynth, nistal2021comparing} used a ProgressiveGAN for controlled musical note synthesis conditioned on one-hot vector for pitch. However, such architectures are under-explored for audio textures, in part because it is difficult to label audio texture data correctly and robustly with control parameter values. Moreover, one-hot representation of the conditioning vector is nominal and sparse, which may produce unconvincing interpolations in the parameter space during generation. Continuous-valued or floating point conditioning has its own challenges, particularly if the range of parameter values is not densely sampled in the training set \cite{ding2020ccgan}, but it is more naturally suited to the goal of generation with interpolated values.

In this paper, we propose a data-driven controllable audio texture morphing strategy with the following contributions: (a) a data-driven parameter distillation method for conditioning GAN for controlled audio texture synthesis, (b) a linear interpolation strategy for conditioning parameters that leads to controlled inter- and intra-class morphing of audio textures, (c) a systematic comparison of our method with existing methods through a set of existing and new objective metrics, (d) our code for parameter distillation through an audio classifier 
    and for GAN training.
\vspace{-0.3cm}
\section{Conditional GAN}
\vspace{-0.2cm}
In this work, we identify two types of continuous conditional parameters - \textit{class-identity parameters} $C$ and \textit{intra-class parameters} $P$. Although they function in the same way during GAN training, class-identity parameters are derived from a classifier trained on the same dataset used to train the GAN. Intra-class parameters are the ones related to the semantics within an audio class. For example, \textit{strength} is an intra-class parameter for the audio texture class wind. Class-identity parameters, by construction, have semantics computed from the dataset, and can be used to navigate between classes while intra-class parameters have externally imposed semantics and may or may not correlate with the class labels. 
We explore two strategies of multi-dimensional conditioning with intra-class and class-identity parameters: (1) two one-hot conditioning vectors, one for representing intra-class parameter, and the other for representing class-identity, called \textit{One-hot GAN}
, and (2) multi-dimensional floating point soft-labels extracted from the penultimate layer of a pre-trained audio classifier representing class-identity conditioning parameters, along with a 1-dimensional floating point intra-class conditional parameter, called \textit{MorphGAN}. 


\vspace{-0.2cm}
\subsection{One-Hot GAN}
\vspace{-0.2cm}
We adopt Engel et al.'s\cite{engel2019gansynth} progressive-GAN with one-hot conditioning (Figure \ref{fig:blockdiagram}(a)). The intra-class parameter $P$ has dimension $q$ equal to the number of unique control parameter values. The class-identity parameter $C$ has dimension $r$ equal to the number of sound classes. To encourage the generator to use the conditional information, an \textit{auxiliary classification} (AC-criterion) loss is added to the discriminator that learns to predict the conditional vector. The AC-criterion calculates the categorical cross entropy loss between the ground-truth conditional vector and the predicted conditional vector through the discriminator. 

\vspace{-0.3cm}
\subsection{MorphGAN}
\vspace{-0.2cm}
Previously, DarkGAN \cite{nistal2021darkgan} took a knowledge distillation approach and used the probabilities extracted from the output layer of an audio classifier that was trained on an external dataset (AudioSet) as a conditional vector for their GAN. However, labels determined by an external dataset may have little relevance for a specific sound model training set. This can lead to a lack of interpretable control over the generated audio. In \mbox{MorphGAN}, we extract soft-labels from the penultimate layer of an audio classifier that is trained on the sound model training dataset.
Since these multi-dimensional soft-labels are learnt by the classifier, they capture multiple class-related aspects of the sound set. 
These dimensions enable interpretable control over interpolation between points in the latent space of the GAN, generating cyclostationary morphs (a sequence of audio segments each produced with different parameters) across novel in-between sounds.

\begin{figure}
\vspace{-15pt}
    \centering
\scalebox{0.8}{
    \includegraphics[width=\columnwidth]{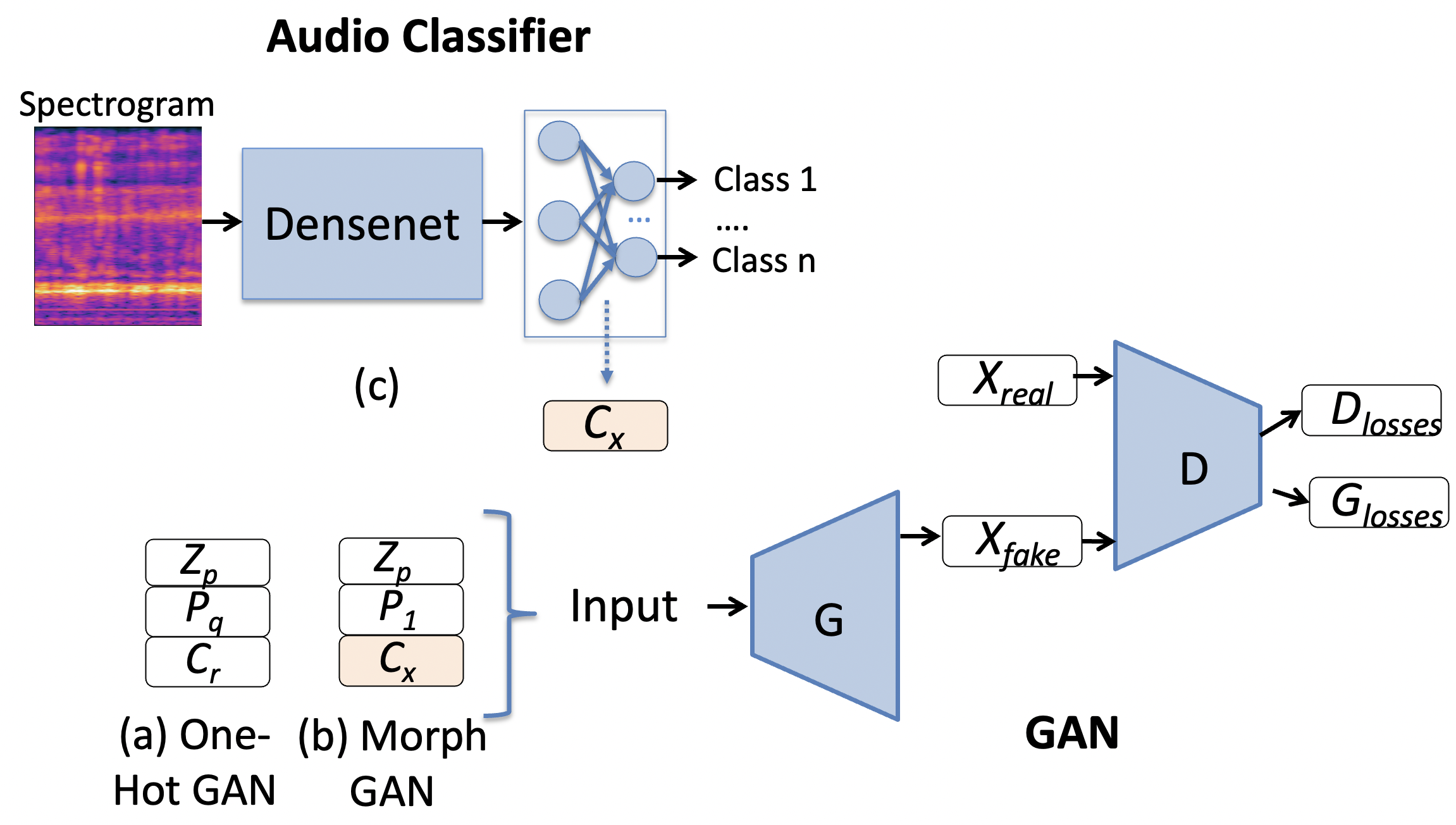}}\vspace{-0.3cm}
    \caption{System overview. GAN input features are a random noise latent vector $Z_p$ ($p$-dim), along with either (a) One-hot vectors for intra-class parameter $P_q$ ($q$-dim) and class-identity parameter $C_r$ ($r$-dim), or (b) Morph-GAN with one dimensional intra-class parameter $P_1$ but $x$ dimensional soft labels for class parameter $C_x$ from the output of the penultimate layer of a pre-trained $n$-class audio classifier.}
    \vspace{-0.3cm}
    \label{fig:blockdiagram}
\end{figure}

MorphGAN uses a single dimensional floating point value for intra-class parameter $P$ and $x$ dimensional floating point soft-labels from the output of the penultimate layer of the audio classifier as the class-identity parameter $C$. Since each data point can have non-exclusive values for each dimension, we use binary cross entropy for auxiliary classifier loss, in which the loss is the sum of the individual binary cross-entropy computation on each dimension,
\vspace{-0.3cm}
\begin{equation}
L=-\frac{1}{K}\sum_{k=1}^{K}[y_k\log(\hat{y}_k)+(1-y_k)\log(1-\hat{y}_k)]
\vspace{-0.3cm}
\label{eq:1}
\end{equation}
where $y_k$ is target conditioning value in the range of [0,1], and $\hat{y}_k$ is the predicted value from the auxiliary classifier for the  $k$th dimension. Subsequently, a sigmoid activation squashes the values in the range [0,1]. Figure \ref{fig:blockdiagram}(c) shows an overview of MorphGAN.

\vspace{-0.3cm}
\section{Experimental Setup}
\vspace{-0.3cm}
\subsection{Datasets}
\vspace{-0.2cm}
We use two types of audio textures in this paper - water, and wind.\\ 
\textbf{Water}: The water sound was recorded by filling a metallic bucket with water at an approximately constant rate over a duration of 30 seconds. We collected 50 audio recordings of different lengths to capture the variation between multiple fillings.The transient sounds at the beginning and end of each sound was trimmed, and then sound was divided into 11 equally spaced time points used as the starting point of a 2-second excerpt labeled with one of 11 different ``fill levels'' normalized to steps of 0.1 in [0,1]. Ten variations were produced from different recordings.\\  
\textbf{Wind}: The wind sound is from the Syntex collection of synthetic datasets \cite{Wyse2022Syntex}. This texture is generated with noise passed through filters modulated with simplex noise\footnote{Section 1.2 of \url{https://animatedsound.com/ismir2022/metrics/appendix_dataset/index.html}}. A ``strength'' parameter controls the wind gust fluctuations defined by the bandpass filter center frequencies. The strength parameter ranges in [0,1] across steps of 0.1, and 10 variations are generated from different random seeds, resulting in 110 audio files each of 2 seconds duration.
\vspace{-0.3cm}
\subsection{Architectures}
\vspace{-0.2cm}
\textbf{GAN:} We adapt the Nistal et al.~\cite{nistal2021comparing} progressive-GAN implementation where generator $G$ transforms the 1D input vector ($Z$+$P$+$C$) to the generated output signal over 5 progressively-grown stages and upsampling CNN blocks. The $Z$ vector is 32-dimensional following \cite{nistal2021darkgan}. The $P$ and $C$ vector dimensions are different across the two GAN variants we employ (Section \ref{sec:exp1}). We found that training models for 120K iterations on batch-size 12 with 20k iterations for the first three stages and batches of 8 files with 30k iterations for the last two produces high quality output (Table \ref{tab:audioquality}).
The audio representation is a magnitude spectrogram computed using the Gabor transform (window size=256, hop size=128). Inversion of the estimated spectrogram is done using phase gradient heap estimation (PGHI) \cite{pruuvsa2017noniterative}. PGHI is a non-iterative phase reconstruction algorithm that uses the mathematical relationship between the magnitude of Gaussian windowed STFT and the phase derivatives in time and frequency of the Fourier transform to reconstruct the phase using only the magnitude spectrogram. Gupta et al.~\cite{gupta2021signal} showed through listening tests that training the GANSynth architecture using only log-mag representation and PGHI inversion produces significantly better audio quality for wideband, non-pitched or fast changing signals. Since the audio data we use in this paper consist of such sounds, eg. water-filling, we used PGHI for reconstruction as it gives better audio quality.
%
%
%
\\
\textbf{Audio Classifier:} The DenseNet model~\cite{palanisamy2020rethinking} pretrained on ImageNet \cite{deng2009imagenet} and fine-tuned for a specific audio dataset can achieve state-of-the-art results for audio classification. We adopted this method for audio classification to generate class soft-labels for MorphGAN. We use the pre-trained Dense Convolutional Network (DenseNet201 PyTorch library), that connects each layer to every other layer in a feed-forward fashion. DenseNet expects a 3-channel input, so a three-channel mel-spectrogram of the audio input is computed using different window sizes and hop lengths of [25ms, 10ms], [50ms, 25ms], and [100ms, 50ms] on each of the channels respectively. The different window sizes and hop lengths ensure the network has different levels of information from the frequency and time domain on each channel, which was shown to perform well for audio classification \cite{palanisamy2020rethinking}. The DenseNet gives a 1,920 dimensional output after which we add two linear layers sequentially of $x$ and $n$ dimensions respectively, where $x$ is a selectable number of soft-labels and $n$ is the number of audio classes. Subsequently, a sigmoid activation function squashes $x$ values between 0 and 1 before using them as class conditional inputs for MorphGAN. 


\vspace{-0.3cm}
\subsection{Models}
\label{sec:exp1}
\begin{figure}[t]
        \centering
\scalebox{0.8}{
        \includegraphics[width=0.6\columnwidth]{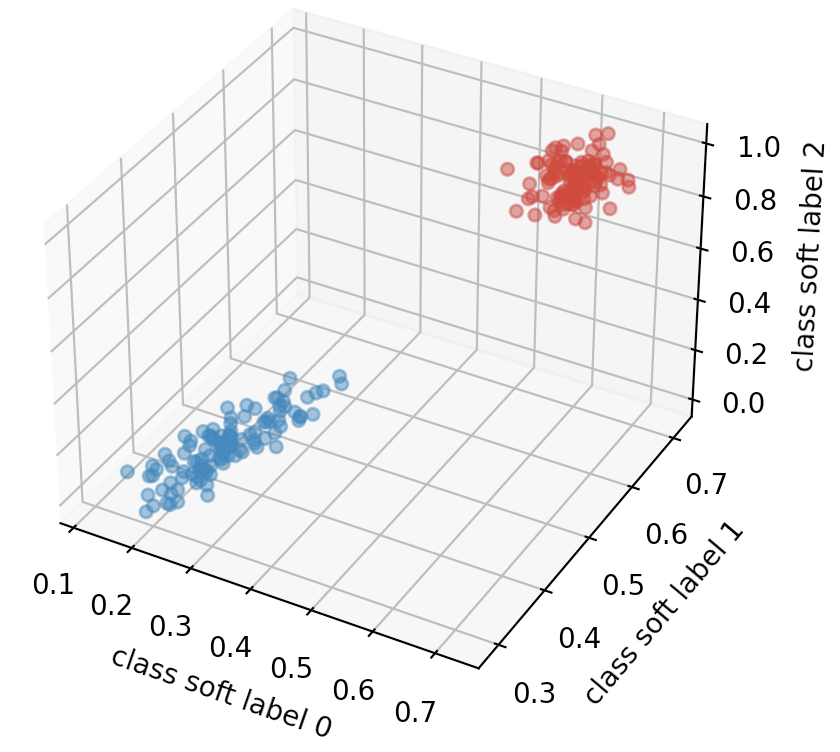}}
        \vspace{-0.3cm}
        \caption{Three dimensional soft label values from the penultimate layer of the audio classifier. These are subsequently used for conditioning MorphGAN. The blue markers are water-filling sounds and the red markers are the wind sounds.}
        \vspace{-0.3cm}
        \label{fig:3d}
        \vspace{-0.2cm}
    \end{figure}

\vspace{-0.2cm}
\textbf{One-Hot GAN}: $P$ is 11 dimensional to represent the 11 discrete values (10 equally-spaced intervals across the range) of the control parameter for the two textures, and $C$ is 2 dimensional. Note that $P$ is a dual serving intra-class parameter, representing fill level for water and strength for wind.\\
\textbf{MorphGAN}: $P$ is a 1D floating point dual serving intra-class parameter for the two textures, discretized to 11 values in [0,1], while $C$ is a 3 dimensional soft-label extracted from the penultimate layer of the audio classifier, values between [0,1]. An 80/20\% split was used for training and validation (val accuracy=100\%). The entire water-wind dataset was then passed through this trained classifier to extract the soft-labels from the penultimate layer. Figure \ref{fig:3d} shows the 3 dimensional soft labels that were learnt by the audio classifier, color-coded with the audio texture class. It is evident that this 3D vector has a wide range of values while also being able to represent the two classes, thus we hypothesize that this 3D vector, when used for class conditioning MorphGAN, will offer more flexibility and control for inter-class morphing than using 1D class vectors.\\
\textbf{Baseline:} As a baseline for comparison, we use Zynaptic's MORPH2.0\footnote{https://www.zynaptiq.com/morph/morph-overview/}, a commercial real-time plug-in for structural audio morphing. We chose their ``classic'' (vocoder-like) interpolation algorithm that uses signal processing to model the timbral shape for every time frame of the two audio inputs, and then interpolates between these models, transforming one sound into the other. Other open-source toolkits, such as sound morphing toolbox \cite{caetano2019morphing} fail for non-pitched audio textures such as water-filling as they employ matching of harmonics in the sound, thus restricting their use in our experiments.
 

We explore two kinds of morphs, inter-class morphs, that interpolate between two points in class-identity conditioning dimensions, and intra-class morphs, that interpolate between two points in intra-class conditioning dimensions.
In the Morph2.0 baseline, we compute interpolations between two sounds from the original data, where for inter-class interpolations, the two sounds belong to water and wind classes, and for intra-class interpolations, the two sounds belong to the same class but two extreme intra-class parameter (fill-level/strength) values. The interpolation is along the morph axis (cross-fade axis=0) of the interface.
\vspace{-0.3cm}
\section{Results}
\vspace{-0.3cm}
Audio examples and the evaluation code to generate the metrics are available on our webpage. 
\footnote{https://animatedsound.com/research/morphgan\_icassp2023/}.
\vspace{-0.4cm}
\subsection{Audio Quality}
\vspace{-0.2cm}
We use the Fréchet Audio Distance (FAD)~\cite{kilgour2019frechet} metric (distance between the distributions of the embeddings of real and synthesized audio data extracted from a pre-trained VGGish model) to evaluate audio synthesis quality as it has been shown to be consistent with human judgements \cite{kilgour2019frechet, nistal2021comparing, gupta2021signal}. 
We compute the FAD for the wind and water sounds generated by the two GAN models, as well as the original one-hot model with latent vector size and training iterations same as in \cite{nistal2021comparing,gupta2021signal}, as shown in Table \ref{tab:audioquality}. We use the training data as the reference distribution and generate 10 variations per condition from each GAN as the test distributions. Our one-hot GAN has reduced dimensions for $Z$ and fewer training iterations but shows similar performance as the original one-hot model. This lightweight architecture has the advantage of reduced training time, and is more suitable for the limited but targeted range of sounds in our dataset. Overall, MorphGAN performs better than the One-Hot GANs. 

\begin{table}[h]
    \centering
    \small
    \vspace{-0.5cm}
    \caption{FAD between generated distribution and real distribution. $\downarrow$ indicates smaller is better.}
    \scalebox{0.65}{
    \begin{tabular}{|p{0.27\columnwidth}|p{0.40\columnwidth}|p{0.14\columnwidth}|p{0.14\columnwidth}|}
    \hline
         \textbf{Architecture} &\textbf{Details}&\textbf{FAD-water($\downarrow$)}&\textbf{FAD-wind($\downarrow$)}\\\hline
         Original one-hot \cite{nistal2021comparing,gupta2021signal}& 128-D $Z$, one-hot $P$,$C$, 1.2M training iterations&5.83&1.25\\\hline
         One-Hot GAN (Reduced $Z$ dims)& 32-D $Z$, one-hot $P$,$C$, 120K training iterations&5.04&1.21\\\hline
         MorphGAN&32-D $Z$, 1-D FP $P$, 3-D FP $C$, 120K training iterations&\textbf{3.13}&\textbf{0.87}\\\hline
    \end{tabular}}
    \label{tab:audioquality}
    \vspace{-0.4cm}
\end{table}

\vspace{-0.3cm}
\subsection{Intra-class Morphing}\label{sec:results-intra-class-morphing}
\vspace{-0.2cm}

We quantify interpolation smoothness of the intra-class morphed sounds by adopting the parameter sensitivity metric from~\cite{gupta2022parameter}. 
This sensitivity metric evaluates the linearity of change in the perceptual distance of an interpolated sound as the intra-class control parameter $P$ is varied from its lowest to its highest value linearly. This linearity of change is quantified using the Pearson's correlation coefficient. Amongst the perceptual distance measures discussed in \cite{gupta2022parameter}, we use Gram Matrix (GM) loss and FAD because  these measures showed high correlation with human perception for the audio textures in this study. Table \ref{tab:result-intra-class-morph-paramsense} shows that MorphGAN is able to produce more perceptually linear intra-class morphs than One-Hot GAN and Morph2.

\begin{table}[h]
\begin{center}
\vspace{-0.5cm}
\caption{Intra-class morphing. $\uparrow$ indicates larger values are better.}
\scalebox{0.65}{
    \begin{tabular}{|p{0.25\columnwidth}|p{0.25\columnwidth}|p{0.25\columnwidth}|p{0.25\columnwidth}|}
        \hline
        \multicolumn{2}{|c|}{\multirow{2}{*}{ \textbf{Architecture}}}&\multicolumn{2}{c|}{\textbf{Control Parameter Sensitivity($\uparrow$)}}\\\cline{3-4}
        \multicolumn{2}{|c|}{}& \textbf{w/ GM Loss}&{ \textbf{w/ FAD}}\\
        \hline
        \multirow{2}{*}{ \textbf{MorphGAN}}&Wind&\textbf{0.97}&\textbf{0.98}\\\cline{2-4}
        &Water&\textbf{0.99}&\textbf{0.95}\\\cline{1-4}
        \multirow{2}{*}{ \textbf{One-Hot GAN}}&Wind&0.80&0.75\\\cline{2-4}
        &Water&0.47&0.74\\\cline{1-4}
        \multirow{2}{*}{ \textbf{Morph2}}&Wind&0.77&0.19\\\cline{2-4}
        &Water&0.90&0.64\\\cline{1-4}
        \hline
    \end{tabular}
}
\label{tab:result-intra-class-morph-paramsense}
\vspace{-0.5cm}
\end{center}
\end{table}

\vspace{-0.5cm}
\subsection{Inter-class Morphing}\label{sec:inter-class-morphing-results}
\vspace{-0.2cm}
To evaluate inter-class morphing, we analyse the effectiveness of the algorithms to (1) linearly/smoothly morph between classes, and (2) their ability to generalize to out-of-distribution (OoD) points in the class parameter space $C$.

For morph smoothness, we adapt the parameter sensitivity metric outlined in the previous section by measuring the GM Loss and FAD between class interpolated samples. Table \ref{tab:results-inter-class-morph} shows that MorphGAN is able to produce smoother linear interpolations between classes than the other methods.

To the best of our knowledge, there is no standard evaluation technique to test OoD generalizability during morphing. We thus develop two additional metrics: \textbf{\textit{Distribution Closeness}} and \textbf{\textit{Distribution Centeredness}} of the samples generated using OoD class parameter values in comparison with the training data.
Specifically, we choose the out-of-distribution value $k=0.5$ for the three class dimensions of MorphGAN (center of the cube in Figure \ref{fig:3d}), for the two class dimensions of One-Hot GAN, and for the morph axis (center) for Morph2.0.
For each algorithm we measure the FAD between the distribution of samples generated from the center point $k$ and the distribution of samples from each of the two classes. We term the mean of the two distances as \textit{Distribution Closeness} to indicate the algorithms' ability to produce sounds related to the training data at this OoD point. 
Further, we refer to the difference between the two distances as \textit{Distribution Centeredness} to indicate the skew of the center point towards any one class.  Table \ref{tab:results-inter-class-morph} shows that the OoD center point of the class parameter space $C$ of MorphGAN is perceptually closer and centered between both the classes and thus can generate more perceptually meaningful and novel morphs in the neighborhood of that location compared to One-Hot or Morph2.


\begin{table}[h]
\begin{center}
\small
\vspace{-0.5cm}
\caption{Inter-class morphing. $\uparrow$ indicates larger values are better.}
\scalebox{0.6}{
    \begin{tabular}{|p{0.20\columnwidth}|p{0.20\columnwidth}|p{0.20\columnwidth}|p{0.17\columnwidth}|p{0.23\columnwidth}|}
        \hline
        \multirow{3}{0.20\columnwidth}{ \textbf{Architecture}}&\multicolumn{2}{c|}{\textbf{Class Parameter Sensitivity($\uparrow$)}}&\multirow{3}{0.17\columnwidth}{\textbf{Distribution Closeness($\downarrow$)}}&\multirow{3}{0.23\columnwidth}{\textbf{Distribution Centeredness($\downarrow$)}}\\
        &\multicolumn{2}{c|}{}&&\\\cline{2-3}
        &\textbf{w/ GM Loss}&{\textbf{w/ FAD}}&&\\
        \hline
        \textbf{MorphGAN}&\textbf{0.96}&\textbf{0.90}&\textbf{8.03}&\textbf{6.00}\\
        \hline
        \textbf{One-Hot GAN}&\textbf{0.96}&0.76&14.05&13.70\\
        \hline
        \textbf{Morph2}&0.84&0.25&16.15&8.70\\
        \hline
    \end{tabular}
}
\label{tab:results-inter-class-morph}
\vspace{-0.7cm}
\end{center}
\end{table}

\begin{figure}[b]
\centering
\vspace{-0.5cm}
\scalebox{0.7}{
\minipage{0.49\columnwidth}
  \includegraphics[width=\linewidth]{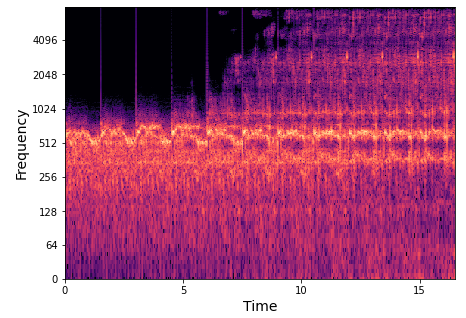}
\endminipage\hfill
\minipage{0.49\columnwidth}
  \includegraphics[width=\linewidth]{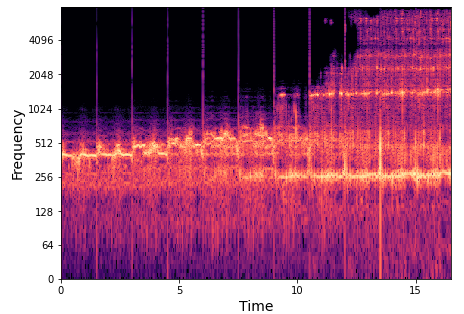}
\endminipage\hfill}
\vspace{-0.2cm}
\caption{Concatenated 2s audio outputs from (a) One-Hot GAN, and (b) MorphGAN as the class parameter $C$ interpolates between values for wind to water in 11 steps while keeping $P$ fixed. }
\label{fig:classinterp}
\vspace{-15pt}
\end{figure}

Qualitatively, the One-Hot morph samples exhibit a stickiness towards one class, and towards the center of the interpolation, there is a sudden transition to the second class, resulting in an abrupt interpolation.
In MorphGAN interpolations, the frequency components of the wind class gradually modify and merge with the frequency components of the water class which corresponds to the perception of a smooth morph (Figure \ref{fig:classinterp} and webpage). 

\begin{figure}[t]
\vspace{-0.5cm}
\centering
\scalebox{0.8}{
\minipage{0.49\columnwidth}
  \includegraphics[width=\linewidth]{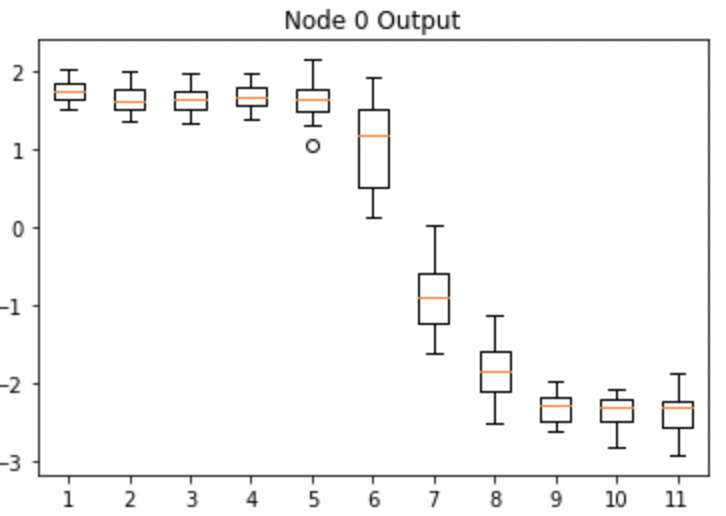}
\endminipage\hfill
\minipage{0.49\columnwidth}
  \includegraphics[width=\linewidth]{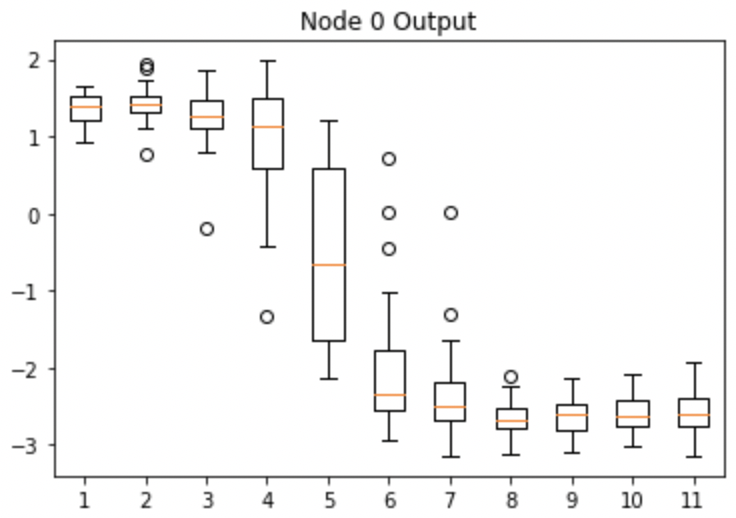}
\endminipage\hfill}
\vspace{-0.2cm}
\caption{Output activation values for node0 (water) of classifier for audio generated from (a) One-Hot GAN, and (b) MorphGAN. The Y-axis is the node0 (water) output from the classifier, and the X-axis is the class parameter interpolated from water to wind.}
\label{fig:consistency}
\vspace{-0.5cm}
\end{figure}

To examine this objectively, we sample a path in the class identity parameter $C$ between a wind and a water sound at 11 points, and for each, we generate 20 audio files for random values of $Z$. The generated audio files are passed back through the classifier and we plot box plots of the output class node0 (water class) values for One-Hot and MorphGAN (Figure \ref{fig:consistency}). Both One-Hot and MorphGAN show consistent outputs towards the class end-points (small std dev in the boxes). However, for class values in between the end points for which neither the classifier nor the GAN were trained, One-Hot shows smaller spread than MorphGAN. This indicates that MorphGAN produces novel morphing sounds with characteristics distinct from the classifier training data whereas the One-Hot tends to stick to one or the other of the two classes. This reinforces our qualitative observation about the same and limits the exploration of sounds in between classes using the One-Hot representation.

\vspace{-0.4cm}
\subsection{Semantic exploration of inter-class morphing}
\begin{figure}[!htb]
\centering
\vspace{-0.2cm}
\scalebox{0.9}{
\minipage{0.32\columnwidth}
  \includegraphics[width=\linewidth]{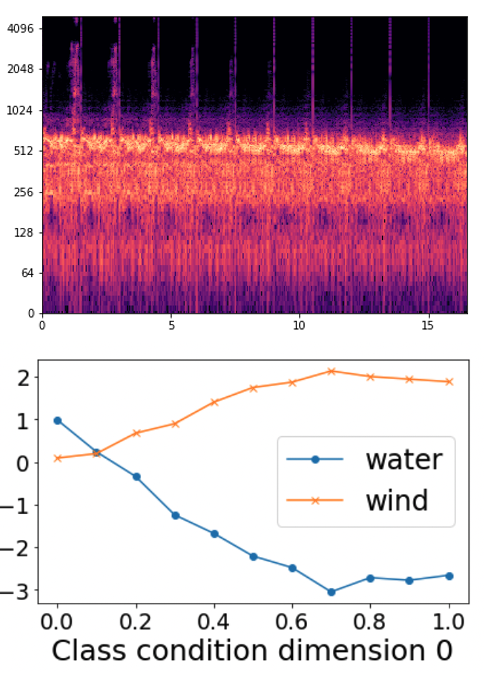}
\endminipage\hfill
\minipage{0.32\columnwidth}
  \includegraphics[width=\linewidth]{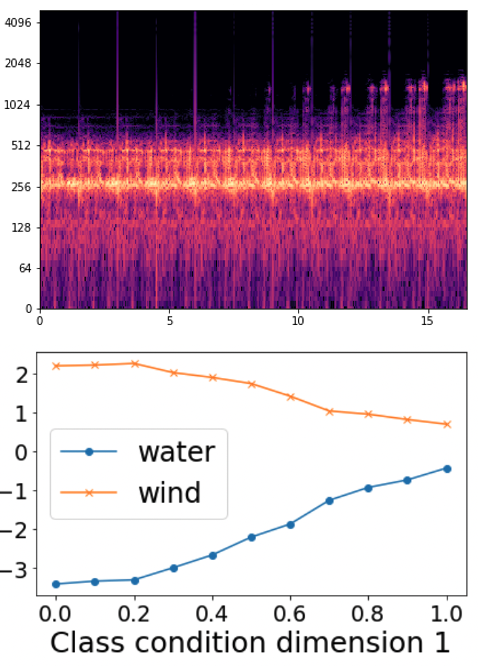}
\endminipage\hfill
\minipage{0.32\columnwidth}%
  \includegraphics[width=\linewidth]{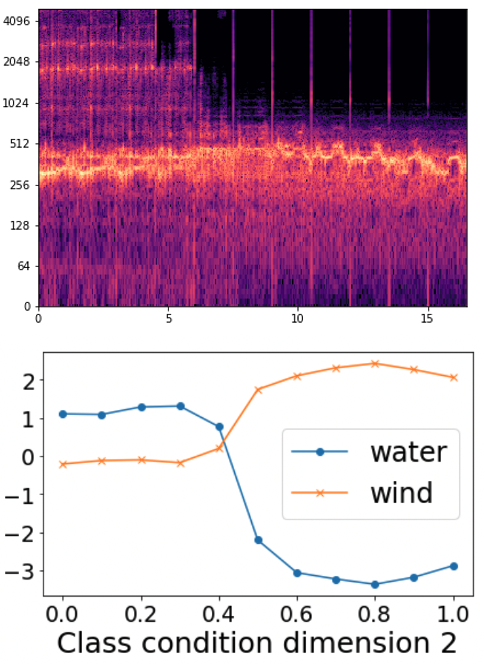}
\endminipage}
\vspace{-0.2cm}
\caption{Spectrogram (top) of concatenated audio outputs of 2s and corresponding audio classifier node activations (bottom) as class parameter (a) dimension 0, (b) dimension 1, and (c) dimension 2 are varied from 0 to 1 at steps of 0.1. Other dimensions are fixed.}
\vspace{-0.3cm}
\label{fig:semantic}
\end{figure}

\vspace{-0.2cm}
To analyse the semantic control of the three class parameter dimensions $C$ of MorphGAN, we varied each dimension from 0 to 1 in steps of 0.1, while keeping all other dimensions constant at 0.5, and fixing a random $Z$ vector (Figure \ref{fig:semantic} (top)). Qualitatively, we can describe variation in the first $C$ dimension 0 as taking the texture from a gurgly-wind sound to a wind-like sound. Lower parameter values also contain higher frequency components from water sounds. Dimension 1 variation takes the texture from a windy whooshing sound to a more watery swish-like sound, where the higher values of this dimension introduce the higher frequency components but at lower amplitudes. Dimension 2 variation moves the texture between water-like and wind-like sounds.

To gain objective insight, we passed these generated audio files back through the audio classifier and plotted the value of the two output node (water and wind) values in Figure \ref{fig:semantic} (bottom). The classification node output values show a trend reinforcing what we hear. For example, dimension 0 variation initially shows a somewhat ambiguous class pattern which changes to predominantly windy. Dimension 1 is smoothly varying, but changes quality while remaining water-like. Dimension 2 (given the values of the other dimension) takes the sound from a fairly clear water to a clear wind sound. These controls afford a perceptual variety of paths between any two endpoints with access to novel sounds that can be explored creatively - for example, to create a wind amplitude pattern modulating a water sound, or to add a bubbly quality to a a wind sound.

\vspace{-0.3cm}
\section{Conclusion}
\vspace{-0.3cm}
%
In this work, we show that class parameters derived from an audio classifier trained on target data is effective for producing convincing morphs between different audio textures. We demonstrate that these class conditional parameters also provide multiple interpretable control dimensions for morphing between two sounds along different paths. We also show that the class parameters consistently produce the intended class across the latent Z space. In the future, improving the consistency of arbitrary control parameters along with wider range of audio textures need to be explored. Future work will also include perceptual listening tests using both audio experts as well a novice listeners and include think-aloud studies to comprehensively analyze the effect of deep learning algorithms over existing baselines. 
This work is a step towards building data-driven controllable audio texture morphing frameworks. 



\balance{}
\bibliographystyle{IEEEbib}
\bibliography{refs.bib}

\begin{thebibliography}{10}

\bibitem{caetano2011morphing}
Marcelo Caetano,
\newblock {\em Morphing isolated quasi-harmonic acoustic musical instrument
  sounds guided by perceptually motivated features},
\newblock Ph.D. thesis, Paris 6, 2011.

\bibitem{kazazis2016sound}
Savvas Kazazis, Philippe Depalle, and Stephen McAdams,
\newblock ``Sound morphing by audio descriptors and parameter interpolation,''
\newblock in {\em Proceedings of the 19th International Conference on Digital
  Audio Effects (DAFx-16). Brno, Czech Republic}, 2016.

\bibitem{slaney1996automatic}
Malcolm Slaney, Michele Covell, and Bud Lassiter,
\newblock ``Automatic audio morphing,''
\newblock in {\em 1996 IEEE International Conference on Acoustics, Speech, and
  Signal Processing Conference Proceedings}. IEEE, 1996, vol.~2, pp.
  1001--1004.

\bibitem{mcdermott2011sound}
Josh~H McDermott and Eero~P Simoncelli,
\newblock ``Sound texture perception via statistics of the auditory periphery:
  evidence from sound synthesis,''
\newblock {\em Neuron}, vol. 71, no. 5, pp. 926--940, 2011.

\bibitem{antognini2019audio}
Joseph~M Antognini, Matt Hoffman, and Ron~J Weiss,
\newblock ``Audio texture synthesis with random neural networks: Improving
  diversity and quality,''
\newblock in {\em International Conference on Acoustics, Speech and Signal
  Processing (ICASSP)}. IEEE, 2019, pp. 3587--3591.

\bibitem{gupta2021signal}
Chitralekha Gupta, Purnima Kamath, and Lonce Wyse,
\newblock ``Signal representations for synthesizing audio textures with
  generative adversarial networks,''
\newblock in {\em Sound and Music Computing (SMC)}, 2021.

\bibitem{choi2022proposal}
Keunwoo Choi, Sangshin Oh, Minsung Kang, and Brian McFee,
\newblock ``A proposal for \protect{F}oley sound synthesis challenge,''
\newblock {\em arXiv preprint arXiv:2207.10760}, 2022.

\bibitem{caetano2019morphing}
Marcelo Caetano,
\newblock ``Morphing musical instrument sounds with the sinusoidal model in the
  sound morphing toolbox,''
\newblock in {\em International Symposium on Computer Music Multidisciplinary
  Research}. Springer, 2019, pp. 481--503.

\bibitem{ezzat2005morphing}
Tony Ezzat, Ethan Meyers, James Glass, and Tomaso Poggio,
\newblock ``Morphing spectral envelopes using audio flow,''
\newblock in {\em Ninth European Conference on Speech Communication and
  Technology}, 2005.

\bibitem{ulyanov2016audio}
Dmitry Ulyanov and Vadim Lebedev,
\newblock ``Audio texture synthesis and style transfer,''
\newblock {\em [Blog post]. Available from: https://dmitryulyanov. github.
  io/audio-texture-synthesis-and-style-transfer/[accessed 16 Jan 2022]}, 2016.

\bibitem{caracalla2020sound}
Hugo Caracalla and Axel Roebel,
\newblock ``Sound texture synthesis using \protect{RI} spectrograms,''
\newblock in {\em ICASSP 2020-2020 IEEE International Conference on Acoustics,
  Speech and Signal Processing (ICASSP)}. IEEE, 2020, pp. 416--420.

\bibitem{gatys2015texture}
Leon Gatys, Alexander~S Ecker, and Matthias Bethge,
\newblock ``Texture synthesis using \protect{C}onvolutional \protect{N}eural
  \protect{N}etworks,''
\newblock {\em Advances in neural information processing systems}, vol. 28, pp.
  262--270, 2015.

\bibitem{luo2019learning}
Yin-Jyun Luo, Kat Agres, and Dorien Herremans,
\newblock ``Learning disentangled representations of timbre and pitch for
  musical instrument sounds using gaussian mixture variational autoencoders,''
\newblock {\em International Society of Music Information Retrieval (ISMIR)},
  2019.

\bibitem{engel2017neural}
Jesse Engel, Cinjon Resnick, Adam Roberts, Sander Dieleman, Mohammad Norouzi,
  Douglas Eck, and Karen Simonyan,
\newblock ``Neural audio synthesis of musical notes with wavenet
  autoencoders,''
\newblock in {\em International Conference on Machine Learning}. PMLR, 2017,
  pp. 1068--1077.

\bibitem{engel2019gansynth}
Jesse Engel, Kumar~Krishna Agrawal, Shuo Chen, Ishaan Gulrajani, Chris Donahue,
  and Adam Roberts,
\newblock ``Gansynth: Adversarial neural audio synthesis,''
\newblock {\em arXiv preprint arXiv:1902.08710}, 2019.

\bibitem{nistal2021comparing}
Javier Nistal, Stefan Lattner, and Gael Richard,
\newblock ``Comparing representations for audio synthesis using generative
  adversarial networks,''
\newblock in {\em 2020 28th European Signal Processing Conference (EUSIPCO)}.
  IEEE, 2021, pp. 161--165.

\bibitem{ding2020ccgan}
Xin Ding, Yongwei Wang, Zuheng Xu, William~J Welch, and Z~Jane Wang,
\newblock ``Ccgan: Continuous conditional generative adversarial networks for
  image generation,''
\newblock in {\em International Conference on Learning Representations}, 2020.

\bibitem{nistal2021darkgan}
Javier Nistal, Stefan Lattner, and Ga{\"e}l Richard,
\newblock ``Darkgan: Exploiting knowledge distillation for comprehensible audio
  synthesis with \protect{GAN}s,''
\newblock {\em arXiv preprint arXiv:2108.01216}, 2021.

\bibitem{Wyse2022Syntex}
Lonce Wyse and Prashanth~Thattai Ravikumar,
\newblock ``Syntex: parametric audio texture datasets for conditional training
  of instrumental interfaces.,''
\newblock {\em International Conference on New Interfaces for Musical
  Expression}, 4 2022,
\newblock https://nime.pubpub.org/pub/0nl57935.

\bibitem{pruuvsa2017noniterative}
Zden{\v{e}}k Pr{\u{u}}{\v{s}}a, Peter Balazs, and Peter~Lempel S{\o}ndergaard,
\newblock ``A noniterative method for reconstruction of phase from
  \protect{STFT} magnitude,''
\newblock {\em IEEE/ACM Transactions on Audio, Speech, and Language
  Processing}, vol. 25, no. 5, pp. 1154--1164, 2017.

\bibitem{palanisamy2020rethinking}
Kamalesh Palanisamy, Dipika Singhania, and Angela Yao,
\newblock ``Rethinking \protect{CNN} models for audio classification,''
\newblock {\em arXiv preprint arXiv:2007.11154}, 2020.

\bibitem{deng2009imagenet}
Jia Deng, Wei Dong, Richard Socher, Li-Jia Li, Kai Li, and Li~Fei-Fei,
\newblock ``Imagenet: A large-scale hierarchical image database,''
\newblock in {\em 2009 IEEE conference on computer vision and pattern
  recognition}. Ieee, 2009, pp. 248--255.

\bibitem{kilgour2019frechet}
Kevin Kilgour, Mauricio Zuluaga, Dominik Roblek, and Matthew Sharifi,
\newblock ``Fr{\'e}chet \protect{A}udio \protect{D}istance: A reference-free
  metric for evaluating music enhancement algorithms.,''
\newblock in {\em INTERSPEECH}, 2019, pp. 2350--2354.

\bibitem{gupta2022parameter}
Chitralekha Gupta, Yize Wei, Zequn Gong, Purnima Kamath, Zhuoyao Li, and Lonce
  Wyse,
\newblock ``Parameter sensitivity of deep-feature based evaluation metrics for
  audio textures,''
\newblock {\em arXiv preprint arXiv:2208.10743}, 2022.

\end{thebibliography}
\label{sec:refs}

\end{document}